\documentclass[aps,pre,preprint,superscriptaddress,amsmath,amssymb,nofootinbib]{revtex4-2}

\usepackage{amsmath}
\usepackage{graphicx}
\usepackage{amssymb}
\usepackage{dcolumn}
\usepackage{bm}
\usepackage{color}
\usepackage{float}
\usepackage{upgreek}
\usepackage{gensymb}
\usepackage[final]{hyperref} 
\hypersetup{
	colorlinks=true,       
	linkcolor=blue,        
	citecolor=blue,        
	filecolor=magenta,     
	urlcolor=blue
}
\usepackage{upgreek}
\usepackage[normalem]{ulem}

\makeatletter

\begin{document}
\title{Diffusion of gravitactic chiral active Brownian particles in an asymmetric channel}
	
\author{Narender Khatri}
\thanks{Corresponding author: narender@mail.jiit.ac.in}
\affiliation{Department of Physics and Material Science and Engineering, Jaypee Institute of Information Technology, Noida, Uttar Pradesh 201304, India}

\author{Vikas Sharma}
\thanks{Email: vikassharma12387@gmail.com}
\affiliation{Department of Computer Science and Engineering, Jaypee Institute of Information Technology, Noida, Uttar Pradesh 201304, India}

\author{Anton F. Burnet}
\thanks{Email: anton.burnet@lmu.de}
\affiliation{Faculty of Physics and Center for NanoScience, Ludwig-Maximilians-Universit\"at M\"unchen, Munich 80752, Germany}
\affiliation{Department of Veterinary Sciences, Ludwig-Maximilians-Universit\"at M\"unchen, Munich 80752, Germany}

\author{Suneet Kumar Awasthi}
\thanks{Corresponding author: suneet\_electronic@yahoo.com}
\affiliation{Department of Physics and Material Science and Engineering, Jaypee Institute of Information Technology, Noida, Uttar Pradesh 201304, India}

\date{\today}
	
\begin{abstract}

The diffusion of micro- and nanoswimmers in a fluid, confined within irregular structures that impose entropic barriers, is often modeled using overdamped active Brownian dynamics, where viscous effects are paramount and inertia is negligible. Here, we numerically investigate the diffusive behavior of chiral self-propelled particles in a two-dimensional asymmetric channel subjected to an external torque arising from a gravitational field. We reveal the emergence of resonant diffusion when the external torque $\omega$ approaches the intrinsic angular velocity $\omega_{0}$ of particles. This resonance manifests as a pronounced accumulation of particles near the upper-left corner of the channel, accompanied by an enhanced peak in the effective diffusion coefficient. In particular, it is observed only for low rotational diffusion rates and does not persist beyond moderate values of $\omega_{0}$. Prominent transport features, such as rectification at low values of $\omega$, a monotonic increase in average velocity with $\omega$, and a nonmonotonic response of transport characteristics (average velocity and effective diffusion coefficient) as a function of the rotational diffusion rate near the resonance point, are explained. Furthermore, we show that the transport characteristics depend strongly on the aspect ratio of the channel. For instance, the enhanced diffusion peak becomes more pronounced with increasing aspect ratio, and the average velocity saturates at higher values for wider bottleneck openings. It is conceivable that these findings have a great potential for developing microfluidic and laboratory-on-a-chip devices for particle separation, targeted drug delivery, and advanced active materials.   

\end{abstract}
	
\maketitle

\newpage

\section{Introduction}

Active particles, both biological and synthetic, also referred to as self-propelled particles, microswimmers, or nanoswimmers, convert free energy from their environments into directed motion under nonequilibrium conditions. Examples of biological active agents include bacteria, sperm cells, and microorganisms~\cite{Berg_Brown@Nature:1972,Woolley@Reproduction:2003,Leptos_et_al@PRL:2009,Lauga_Powers@RPP:2009}, while synthetic counterparts include Janus particles, catalytic microswimmers, magnetic microswimmers, and active colloids~\cite{Zhiyu_et_al@JCP:2021,Gompper_et_al@JPCM:2020,Bechinger_et_al@RMP:2016,Elgeti_et_al@RPP:2015}. The mechanisms underlying such active motion, as well as the dynamical properties of these systems, are diverse and have been extensively studied~\cite{Cates@RPP:2012,Wang2015,Zoettl2016,Bechinger_et_al@RMP:2016,Gompper_et_al@JPCM:2020,Gompper_et_al@JPCM:2025,Elgeti_et_al@RPP:2015}. Substantial experimental work has demonstrated the diverse applications of these microscopic or submicroscopic active agents, including microfluidics~\cite{Woodhouse_Dunkel@NC:2017}, microsurgery~\cite{Vyskocil_et_al@AN:2020}, active cargo transport~\cite{Xu_et_al@Small:2019}, targeted drug delivery~\cite{Tang_et_al@SR:2020}, and numerous others~\cite{Soto_et_al@ACIE:2020,Bunea_et_al@Micromachines:2020}. A notable subclass of active systems comprises chiral active particles, which, in addition to self-propulsion, move in curved trajectories due to either internal asymmetry or external torque~\cite{Liebchen_Levis@PRL:2017,Kummel_et_al@PRL:2013,Wang_et_al@PRL:2024}.

These active particles can respond to external fields, torques, and gradients, giving rise to various forms of taxis including chemotaxis~\cite{Khatri_Kapral@Chaos:2024,Vuijk_et_al@PRL:2021}, magnetotaxis~\cite{Erglis_et_al@BJ:2007,Codutti_et_al@eLife:2022}, gravitaxis~\cite{note:Roberts@2006_2010,Chepizhko_Franosch@PRL2022,Hagen_et_al@NC2014,Rusch_et_al@PRE2024}, rheotaxis~\cite{Mathijssen_et_al@NC:2019}, and viscotaxis~\cite{Liebchen_et_al@PRL:2018}. In particular, negative gravitaxis, where directed movement occurs against the force of gravity, has been reported for microorganisms such as Euglena gracilis~\cite{Lebert_Hader@ASR:1999}, Paramecium~\cite{Hemmersbac_et_al@ASR:2001}, and asymmetric self-propelled colloidal particles~\cite{Chepizhko_Franosch@PRL2022,Rusch_et_al@PRE2024,Hagen_et_al@NC2014}. Both theoretical models and experimental observations have demonstrated that the swimmers align with the gravitational field. Recently, Chepizhko and Franosch~\cite{Chepizhko_Franosch@PRL2022} studied the diffusive behavior of chiral active particles subject to an external torque due to the presence of a gravitational field. In this work, they found a resonant diffusion when the external torque approaches the intrinsic angular drift of the particle. They revealed that the angular equation of a gravitactic particle can be mapped to an overdamped, classical, noisy driven pendulum, and that the resonance originates from an underlying bifurcation of this pendulum.

In many situations encountered in soft condensed matter and biological systems, active particles move in confined environments such as narrow channels, pores, or cavities~\cite{Xiao_et_al@ACS:2019,Perkin_Klein@SM:2013}. The irregular shape of confined geometries restricts the volume of the phase space accessible to the particles, thereby creating entropic barriers that significantly influence their transport characteristics~\cite{Zwanzig@JCP:1992,Khatri_Burada@PRE:2020,Khatri_Burada@PRE:2021}. Additionally, confined structures with spatial ratchet asymmetry generate an entropic ratchet potential that drives active directed transport in the system~\cite{Reichhardt_Reichhardt@ARCMP:2017,Khatri_Kapral@JCP:2023,Ghosh_et_al@PRL:2013}. Notably, the transport characteristics and distribution of gravitactic chiral active particles in confined environments remain unexplored and have not been quantitatively scrutinized. 

In this article, we numerically study the transport characteristics and distribution of chiral self-propelled particles confined within a two-dimensional asymmetric channel, subject to an external torque due to the presence of a gravitational field. We employ a minimal overdamped Langevin model to describe the dynamics of self-propelled particles, with their collision dynamics with the channel walls modeled by sliding-reflecting boundary conditions~\cite{Khatri_Kapral@JCP:2023,Ghosh_et_al@PRL:2013,Reichhardt_Reichhardt@ARCMP:2017}. We emphasize how the external torque influences the diffusive behavior of active particles in the asymmetric channel. The rest of this article is organized as follows: in Sec.~\ref{sec: Model}, we introduce our model for gravitactic chiral active particles in a two-dimensional asymmetric channel. Section~\ref{Sec: Distributions} presents the effects of the external torque on the spatial distribution of particles, while Sec.~\ref{Sec: Transport_Properties} discusses their transport characteristics. Finally, the main conclusions of the article are provided in Sec.~\ref{Sec: Conclusions}. 

\section{Model}
\label{sec: Model}

\begin{figure}[hbt]
\centering
\resizebox{0.85\columnwidth}{!}{%
\includegraphics[scale = 1.0]{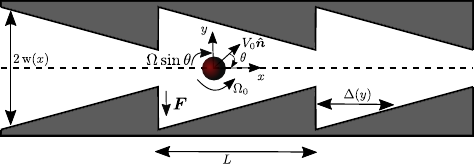}}
\caption{Schematic illustration of an active (self-propelled) Brownian particle diffusing in a two-dimensional triangular channel with periodicity $L$. The self-propelled velocity $V_0 \boldsymbol{\hat{n}}$, angle $\theta$, intrinsic angular velocity $\Omega_{0}$, gravitational force $\boldsymbol{F} = F \boldsymbol{\hat{y}}$, gravitational torque $\Omega \sin \theta $, local width of the channel $2~\mathrm{w}(x)$, and local length of a cell of the channel $\Delta (y)$ are indicated. The particle is restricted from penetrating the rigid channel walls; however, it remains free to rotate and slide along the walls.}
\label{fig:Model}
\end{figure}

Consider a system of active particles dilutely dispersed in a dissipative medium and constrained to diffuse in a two-dimensional asymmetric channel with periodicity $L$ (see Fig.~\ref{fig:Model}). The active particle density is assumed to be sufficiently low such that direct interactions among particles can be ignored. Following established approaches~\cite{Chepizhko_Franosch@PRL2022,Rusch_et_al@PRE2024,Hagen_et_al@NC2014}, we employ a minimal overdamped Langevin model for a particle subjected to an external gravitational field $\boldsymbol{F} = F \boldsymbol{\hat{y}}$. The dynamics of an active particle with position $\boldsymbol{r}  = (x, y)$ and orientation $\boldsymbol{\hat{n}} = (\cos{ \theta }, \sin{\theta })$ is governed by the following coupled Langevin equations    
\begin{equation}\label{Eq:Langevin1}
 \begin{aligned}
    \dot{\bm{r}}(t) &=  V_0 \boldsymbol{\hat{n}} (t) + \sqrt{2 D_t}~ \boldsymbol{\xi} (t), \\
    \dot{\theta}(t) &=  \Omega_{0} - \Omega \, \sin \theta (t) + \sqrt{2 D_r} ~ \zeta (t).
    \end{aligned}
\end{equation}
Here, $V_{0}$ is the self-propelled velocity, $D_t$ and $D_r$ are the translational and rotational diffusion constants, respectively, $\Omega_{0} \, (> 0)$ is the intrinsic angular velocity, and $\Omega \, (> 0)$ is the gravitational torque that is proportional to the external field. Such a set of coupled Langevin equations provides a general framework for modeling a wide class of active systems. 
For systems with thermal noises that satisfy the fluctuation-dissipation relation, $D_t$ and $D_r$ are related to the translational and rotational friction coefficients by the Stokes-Einstein relations: $D_t = k_{B} T/\gamma_t$ and $D_r = k_{B} T/\gamma_r$. However, $\{\gamma_t, \gamma_r, D_t, D_r \}$ are treated as independent parameters for systems subject to athermal noise~\cite{Khatri_Kapral@JCP:2023,Fily_Marchetti@PRL:2012}. The surface coating of the particle is configured in such a way that the external torque aligns the orientation in the horizontal direction perpendicular to the gravitational field, i.e., $\boldsymbol{\hat{n}} = (1, 0)$. The translational and rotational Brownian fluctuations from the surrounding medium are modeled by the random variables $\boldsymbol{\xi} (t)$ and $\zeta (t)$, respectively, having zero mean and unit variances given by $\langle \boldsymbol{\xi} (t) \otimes \boldsymbol{\xi}^{\mathrm{T}} (t')\rangle = \delta (t - t') \boldsymbol{\mathrm{1}}$ and $\langle \zeta (t) \zeta (t') \rangle = \delta (t - t')$, where $\boldsymbol{\mathrm{1}}$ is the unit matrix. 

The two-dimensional asymmetric channel shown in Fig.~\ref{fig:Model} is periodic along the $x$ direction with periodicity $L$ and extends infinitely. The shape of the channel is described by the upper channel wall given by  
\begin{equation}\label{eq:wall}
{\rm w}_u (x) = \begin{cases}
{\rm w}_\mathrm{min}, & x = 0,\\
{\rm w}_\mathrm{max} - ({\rm w}_\mathrm{max} - {\rm w}_\mathrm{min}) \frac{x}{L}, & 0 < x\leq L,
\end{cases}
\end{equation}
where ${\rm w}_\mathrm{min}$ and ${\rm w}_\mathrm{max}$ denote its minimum and maximum half-widths, respectively, ${\rm w}_l (x) = - {\rm w}_u (x)$ represents its lower channel wall, and $2 \, {\rm w} (x) = {\rm w}_u (x) - {\rm w}_l (x)$ corresponds to its local width. The aspect ratio of the channel is defined as $\epsilon = {\rm w}_\mathrm{min}/{\rm w}_\mathrm{max}$, with ${\rm w}_\mathrm{max} = 1$ throughout this work. The local length of a cell of the channel, at a given $y$, is expressed as $\Delta (y) = ({\rm w}_\mathrm{max} - |y|)L/({\rm w}_\mathrm{max} - {\rm w}_\mathrm{min})$ for ${\rm w}_\mathrm{min} < |y| \le {\rm w}_\mathrm{max}$ and takes the constant value $\Delta (y) = L$ for $-{\rm w}_\mathrm{min} \le y  \le {\rm w}_\mathrm{min}$. To ensure the confinement of particles inside the channel, sliding-reflecting boundary conditions are implemented at the channel walls~\cite{Khatri_Kapral@JCP:2023,Ghosh_et_al@PRL:2013,Reichhardt_Reichhardt@ARCMP:2017, Khatri_Burada@PRE:2022}. More precisely, when the particle encounters the channel wall, its translational velocity $\dot{\bm{r}}$ is elastically reflected, while its orientation $\boldsymbol{\hat{n}}$ remains unaffected. As a result, only the velocity component normal to the channel wall is reversed, whereas the tangential component remains unchanged. Consequently, the particle slides along the channel wall until orientational fluctuations redirect it back into the interior of the channel. 

It is convenient to use a dimensionless description~\cite{Khatri_Kapral@JCP:2023}, where spatial coordinates are rescaled by the periodicity of the channel $L$, $\boldsymbol{r'} = \boldsymbol{r}/L$, and time is rescaled by the characteristic diffusion time $\tau = L^2/D_t$, $t' = t/\tau$. In dimensionless form, the set of coupled Langevin equations~(\ref{Eq:Langevin1}) reads
\begin{equation}\label{Eq:Langevin2}
 \begin{aligned}
    \dot{\bm{r}}(t) &=  v_0 \boldsymbol{\hat{n}} (t) + \sqrt{2}~ \boldsymbol{\xi} (t), \\
    \dot{\theta}(t) &=  \omega_{0} - \omega \, \sin \theta (t) + \sqrt{2 \alpha} ~ \zeta (t),
    \end{aligned}
\end{equation} 
where we have dropped the primes for better readability. The dimensionless parameters are given as follows: $v_{0} = V_{0} L/D_{t}$, $\omega_{0} = \Omega_{0} \tau$, $\omega = \Omega \tau$, and $\alpha = D_r \tau$. 
In the absence of the external field, the dynamics reduces to that of a chiral active particle, whose trajectory approximately combines a circular arc of radius $R_{\omega_{0}} = v_{0}/\omega_{0}$ and a persistence length $l_{p} = v_{0}/\alpha$~\cite{Bechinger_et_al@RMP:2016,Teeffelen_Lowen@PRE:2008}, while for $\omega > 0$, the angular motion corresponds to Brownian motion in a tilted washboard potential given by $U(\theta) = -\omega_{0} \theta - \omega \cos{ \theta }$~\cite{Reimann_et_al@PRL:2001,Reimann_et_al@PRE:2002}. Notably, setting $\alpha = 0$ in Eq.~(\ref{Eq:Langevin2}) yields a deterministic angular dynamics equivalent to that of an overdamped driven pendulum, which exhibits a saddle-node bifurcation at the critical value $\omega_{c} = \omega_{0}$~\cite{Strogatz@Book:2018,Sprott@Book:2010,Chepizhko_Franosch@PRL2022}. In analogy with the overdamped driven pendulum and assuming no orientational noise ($\alpha = 0$), a stable fix point arises at $\theta^{*} = \sin^{-1}(\omega_0/\omega)$ for $\omega > \omega_0$, corresponding to a state where the orientational angle is locked within $0 < \theta^{*} < \pi/2$. In contrast, no fix points exist for $\omega < \omega_0$, and the angular motion becomes periodic. We therefore refer to the regimes $\omega > \omega_0$ and $\omega < \omega_0$ as the locked and running states, respectively.

Since the particles are confined in the transverse ($y$) direction, we only calculate their transport characteristics, i.e., the average velocity $v$ and the effective diffusion coefficient $D_{\mathrm{eff}}$, along the longitudinal ($x$) axis. Closed-form expressions for these transport characteristics are not available, as the boundary conditions at the channel walls render the problem of solving the Fokker-Planck equations analytically intractable. Therefore, these transport characteristics are calculated from simulations of the coupled Langevin equations~(\ref{Eq:Langevin2}) in the channel \cite{note:sim}. As an initial condition at time $t = 0$, the particles with randomly assigned orientations are uniformly distributed within a single cell of the channel spanning $x \in [0, 1]$, and the results are obtained by averaging over $10^{4}$ stochastic trajectories. Numerically, $v$ and $D_{\mathrm{eff}}$ along the longitudinal axis are, respectively, calculated as 
\begin{align}
 v &= \lim_{t\to\infty} \frac{\langle x(t) \rangle}{t}, \label{eq:velocity}\\
 D_{\mathrm{eff}}  &= \lim_{t\to\infty} \frac{\langle x^2(t) \rangle - \langle x(t) \rangle ^2 }{2 \,t}. \label{eq:TPs_calculation}
 \end{align}
In the sections that follow, we present results for the average velocity, effective diffusion coefficient, and other relevant properties.

\section{Steady-State Spatial Distribution}\label{Sec: Distributions}

\begin{figure}[H]
\centering
\resizebox{0.82\columnwidth}{!}{%
\includegraphics[scale = 1.0]{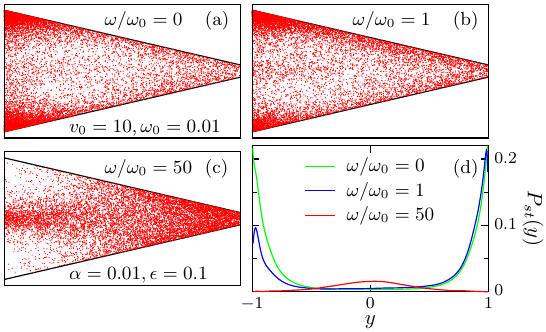}}
\caption{(a)-(c) Steady-state distribution of particles, mapped onto a single cell of the channel, for different values of torque $\omega$. (d) Corresponding probability densities $P_{st} (y)$ along the $y$ direction. The parameters are $v_{0} = 10, \omega_{0} = 0.01, \alpha = 0.01$, and $\epsilon = 0.1$.}
\label{fig:Distribution_Graph1}
\end{figure}

In order to investigate the effect of the external gravitational torque $\omega$ on the diffusion of self-propelled particles in a channel, we first examine their spatial distribution mapped onto a single cell of the channel. Considering $P_{st}(x, y)$ as the steady-state probability density of finding a particle at position $(x, y)$ within the cell, we define the corresponding steady-state probability density at a given $y$ per unit local cell length $\Delta(y)$ as $P_{st}(y)=\int_0^{\Delta(y)} dx \;P_{st}(x,y)/\Delta(y)$. Figure~\ref{fig:Distribution_Graph1} shows the steady-state distribution of particles and the corresponding probability density $P_{st} (y)$ along the $y$ direction for different values of torque $\omega$. In the absence of external torque, i.e., for $\omega/\omega_{0} = 0$, the distribution of particles is inhomogeneous, with most particles accumulating near the left corners of the cell and a slightly higher density above the principal axis of the channel due to $\omega_{0} > 0$. The distribution of particles remains asymmetric with respect to the principal axis of the channel until the chiral radius $R_{\omega_{0}} = v_{0}/\omega_{0}$ is comparable to or greater than the local width of the channel, as reported in Refs.~\cite{Khatri_Burada@PRE:2022,Xue@PhD_Thesis:2014}. The observed inhomogeneous distribution of particles arises from the active motion that breaks detailed balance, combined with the spatial asymmetry imposed by the channel structure~\cite{Ghosh_et_al@PRL:2013, Reichhardt_Reichhardt@ARCMP:2017, Khatri_Kapral@JCP:2023}.  It is interesting to see that near the classical bifurcation point ($\omega/\omega_{0} = 1$), the accumulation of particles becomes much higher near the upper-left corner of the channel compared to near the lower-left corner (see the $P_{st} (y)$ plot for the curve $\omega/\omega_{0} = 1$). Note that preferential accumulation results from the stable fix point at $\theta^{*} \approx \pi/2$ when $\omega/\omega_{0}$ is slightly greater than unity, together with the shape of the channel. This observation provides evidence for the emergence of resonant diffusion close to the bifurcation point~\cite{Chepizhko_Franosch@PRL2022} and for enhanced rectification~\cite{Khatri_Kapral@JCP:2023}. It is worth noting that L\"owen and co-workers have reported the accumulation of self-propelled particles near the corners of microwedge structures and along curved walls, without requiring alignment with external fields~\cite{Bechinger_et_al@RMP:2016,Kaiser_et_al@PRE:2013,Smallenburg_Lowen@PRE:2015}. In particular, when $\omega$ dominates over $\omega_{0}$, particles tend to focus in the middle and at the exit of the channel, and the distribution becomes symmetric about the principal axis of the channel [see Fig.~\ref{fig:Distribution_Graph1}(c)]. This is because, in the regime $\omega \gg \omega_0$, the particle orientations increasingly align with the horizontal direction as $\omega$ increases, effectively propelling them in the forward direction~\cite{Chepizhko_Franosch@PRL2022}.

\begin{figure}[htb!]
\centering
\resizebox{0.82\columnwidth}{!}{%
\includegraphics[scale = 1.0]{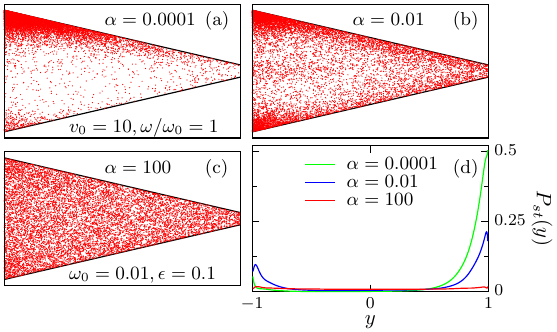}}
\caption{(a)-(c) Steady-state distribution of particles for different values of the scaled rotational diffusion rate $\alpha$ at $\omega/\omega_{0} = 1$. (d) Corresponding probability densities $P_{st} (y)$ versus $y$. The parameters are $v_{0} = 10, \omega_{0} = 0.01, \omega = 0.01$, and $\epsilon = 0.1$.}
\label{fig:Distribution_Graph2}
\end{figure}

The steady-state distribution of particles and the corresponding probability density $P_{st} (y)$ along the $y$ direction are shown in Fig.~\ref{fig:Distribution_Graph2} for different values of the rotational diffusion rate $\alpha$ at the bifurcation point ($\omega/\omega_{0} = 1$). For $\alpha \to 0$, particles predominantly accumulate near the upper-left corner of the cell due to resonant diffusion at $\omega/\omega_{0} = 1$, while their density remains markedly low below the principal axis of the channel. With a further increase in $\alpha$, though still within the small $\alpha$ regime, an additional accumulation of particles develops near the lower-left corner of the cell, as shown in Fig.~\ref{fig:Distribution_Graph2}(b); however, this accumulation remains weaker than that near the upper-left corner. It should be noted that for moderate and large values of $\alpha$, the distribution of particles becomes symmetric about the principal axis of the channel, clearly indicating the complete disappearance of resonant diffusion observed at small $\alpha$. In particular, as $\alpha \to \infty$ where reorientation is rapid, the self-propelled dynamics effectively reduces to passive Brownian motion, resulting in a uniform distribution of particles inside the periodic cell consistent with thermodynamic equilibrium~\cite{Khatri_Burada@JCP:2019,Jacobs@Book:1967,Zwanzig@JCP:1992,Reguera_Rubi@PRE:2001,Kalinay_Percus@PRE:2006} [see Fig.~\ref{fig:Distribution_Graph2}(c)].

\section{Average velocity and effective diffusion}\label{Sec: Transport_Properties}

\begin{figure}[htb!]
\centering
\resizebox{0.55\columnwidth}{!}{%
\includegraphics[scale = 1.0]{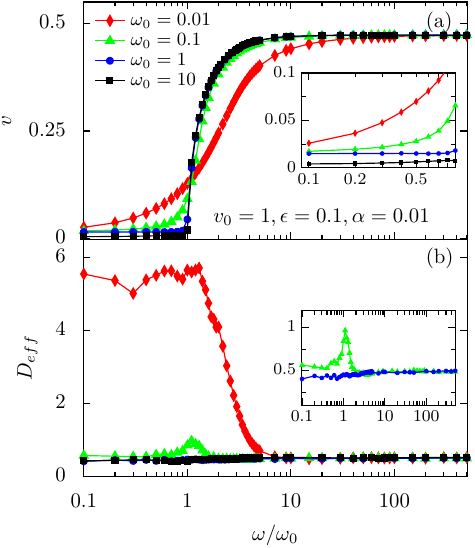}}
\caption{Average velocity $v$ as a function of the torque $\omega$ is shown in (a) for different values of the intrinsic angular velocity $\omega_{0}$. The corresponding effective diffusion coefficient $D_{\mathrm{eff}}$ is shown in (b). The insets plot $v$ and $D_{\mathrm{eff}}$ versus $\omega$ on an expanded scale for selected values of $\omega_0$. Here and below, solid lines serve as guides to the eye. The other parameters are $v_{0} = 1, \alpha = 0.01$, and $\epsilon = 0.1$.}
\label{fig:Transport_Properties_Graph1}
\end{figure}

Figure~\ref{fig:Transport_Properties_Graph1} depicts the dependence of the average velocity $v$ and effective diffusion coefficient $D_{\mathrm{eff}}$ on the torque $\omega$ for different values of the intrinsic angular velocity $\omega_{0}$. For $\omega \ll \omega_{0}$, the external torque can be neglected, and the particles exhibit rectification ($v \neq 0$) in the positive $x$ direction due to the chosen shape of the channel. As $\omega_{0}$ increases, the chiral radius $R_{\omega_{0}} = v_{0}/\omega_{0}$ decreases until the particles can perform closed circular orbits within the compartment without touching the channel walls, resulting in $v \to 0$ [see the inset of Fig.~\ref{fig:Transport_Properties_Graph1}(a)]. 
In the locked state ($\omega > \omega_{0}$), $v$ increases monotonically with $\omega$, eventually reaching a saturation plateau for $\omega \gg \omega_{0}$. This is because in this regime, the particle orientations increasingly align with the horizontal direction, as the gravitational field enforces orientations in this direction (see Fig.~\ref{fig:Distribution_Graph1}). As expected, the saturated value of $v$ is independent of $\omega_{0}$. Note that the convergence of $v$ to its saturated value occurs faster as $\omega_{0}$ increases, indicating that the particle orientations are better locked with increasing $\omega_{0}$. Interestingly, $D_{\mathrm{eff}}$ exhibits an enhanced diffusion peak close to the bifurcation point ($\omega/\omega_{0} = 1$), which is a signature of resonant diffusion. Far above this point, the orientational angle $\theta$ performs small fluctuations most of the time near the minimum of the effective potential $U(\theta)$, keeping $D_{\mathrm{eff}}$ low. As the bifurcation is approached, $U(\theta)$ becomes softer, allowing these fluctuations to become more significant, since the accumulation of particles is much higher near the upper-left corner of the channel compared to the lower-left, thereby enhancing $D_{\mathrm{eff}}$. Below the bifurcation, these fluctuations become too large, leading to a decrease in the accumulation of particles near the upper-left corner of the channel and, consequently, $D_{\mathrm{eff}}$ drops as the harmonic approximation for $U(\theta)$ breaks down due to barrier-crossing events. Therefore, the essence of resonance arises primarily from the enhancement of small fluctuations enabled by the softening of $U(\theta)$.
For larger values of $\omega_{0}$, the enhanced diffusion peak disappears, reflecting that the resonant diffusion vanishes.


\begin{figure}[htb!]
\centering
\resizebox{0.55\columnwidth}{!}{%
\includegraphics[scale = 1.0]{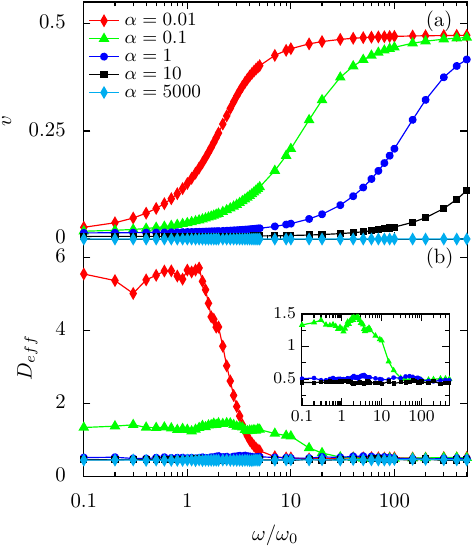}}
\caption{Average velocity $v$ and effective diffusion coefficient $D_{\mathrm{eff}}$ as a function of torque $\omega$ are shown in (a) and (b), respectively, for different values of the rotational diffusion rate $\alpha$. The inset plots $D_{\mathrm{eff}}$ versus $\omega$ for $\alpha = 0.1$, $\alpha = 1$, and $\alpha = 10$ on an expanded scale. The other parameters are $v_{0} = 1, \omega_{0} = 0.01$, and $\epsilon = 0.1$.}
\label{fig:Transport_Properties_Graph2}
\end{figure} 

The variation of $v$ and $D_{\mathrm{eff}}$ with $\omega$ for different values of the rotational diffusion rate $\alpha$ is depicted in Fig.~\ref{fig:Transport_Properties_Graph2}. The qualitative trends for different $\alpha$ remain the same as those described above for various values of $\omega_{0}$. The convergence of $v$, however, to its saturated value becomes slower as $\alpha$ increases because the particle orientations are better locked as $\alpha$ decreases. Additionally, resonant diffusion, manifested as an enhanced diffusion peak, emerges only for low values of $\alpha$. This enhanced diffusion peak vanishes for moderate and high values of $\alpha$ as orientational fluctuations become large in this regime. In the limit $\alpha \to \infty$, where reorientation is rapid, the self-propelled dynamics effectively reduces to passive Brownian motion. Consequently, $v$ approaches zero and $D_{\mathrm{eff}}$ remains low for any value of $\omega$.  

\begin{figure}[htb!]
\centering
\resizebox{0.55\columnwidth}{!}{%
\includegraphics[scale = 1.0]{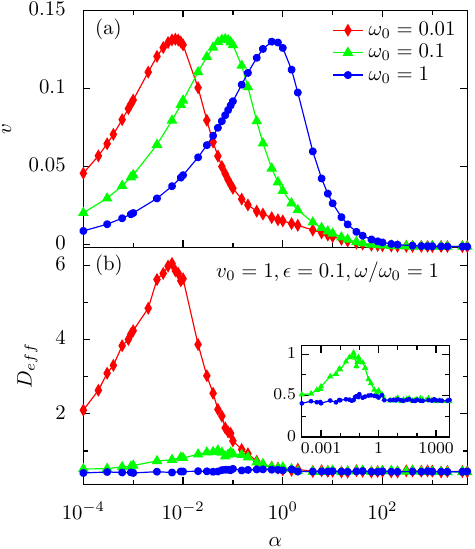}}
\caption{Average velocity $v$ and effective diffusion coefficient $D_{\mathrm{eff}}$ as a function of the rotational diffusion rate $\alpha$ are shown in (a) and (b), respectively, for different values of the intrinsic angular velocity $\omega_{0}$ at $\omega/\omega_{0} = 1$. The inset plots $D_{\mathrm{eff}}$ versus $\alpha$ for $\omega_{0} = 0.1$ and $\omega_{0} = 1$ on an expanded scale. The other parameters are $v_{0} = 1$ and $\epsilon = 0.1$.}
\label{fig:Transport_Properties_Graph3}
\end{figure}

Figure~\ref{fig:Transport_Properties_Graph3} shows $v$ and $D_{\mathrm{eff}}$ versus $\alpha$ for different values of the intrinsic angular velocity $\omega_{0}$ at the bifurcation point $\omega/\omega_{0} = 1$. In the small $\alpha$ limit, an increase in $\alpha$ affects the particle dynamics most prominently near the bifurcation since an additional accumulation of particles develops near the lower-left corner of the channel (see Fig.~\ref{fig:Distribution_Graph2}), leading to an increase in $v$ and $D_{\mathrm{eff}}$ with $\alpha$. In the limit $\alpha \to \infty$, the self-propelled motion approaches passive Brownian motion, resulting in $v \to 0$ and $D_{\mathrm{eff}}$ plateaus at a small value. Therefore, as one might expect, $v$ and $D_{\mathrm{eff}}$ exhibit a peak at an optimal value of $\alpha$. The peak in $D_{\mathrm{eff}}$ is more pronounced as $\omega_{0}$ decreases, while the maximum value of $v$ is unaffected by variations in $\omega_{0}$. As resonant diffusion emerges only for low values of $\alpha$ near the bifurcation, the peak in $D_{\mathrm{eff}}$ appears only in the low $\alpha$ regime and vanishes for larger values of $\omega_{0}$ due to the disappearance of resonance (see the inset of Fig.~\ref{fig:Transport_Properties_Graph3}). It should be noted that the nonmonotonic behavior of $v$ as a function of $\alpha$ is observed only near the bifurcation ($\omega/\omega_{0} \approx 1$); in contrast, in both the locked ($\omega/\omega_{0} > 1$) and running ($\omega/\omega_{0} < 1$) states, $v$ is found to be monotonically decreasing with $\alpha$.

\begin{figure}[htb!]
\centering
\resizebox{0.55\columnwidth}{!}{%
\includegraphics[scale = 1.0]{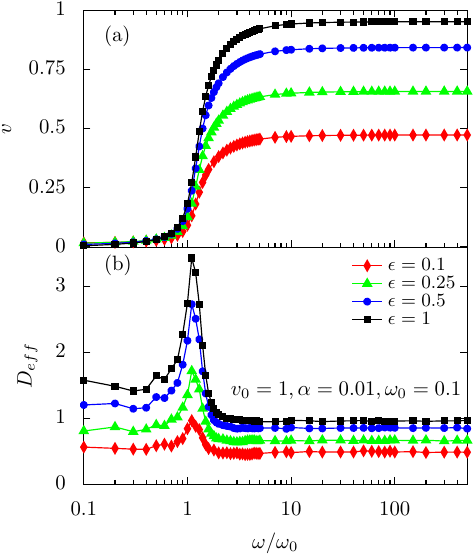}}
\caption{Average velocity $v$ and effective diffusion coefficient $D_{\mathrm{eff}}$ as a function of torque $\omega$ are shown in (a) and (b), respectively, for different values of the aspect ratio of the channel $\epsilon$. The other parameters are $v_{0} = 1$, $\alpha = 0.01$, and $\omega_{0} = 0.1$.}
\label{fig:Transport_Properties_Graph4}
\end{figure}

Figure~\ref{fig:Transport_Properties_Graph4} shows the behavior of $v$ and $D_{\mathrm{eff}}$ as a function of $\omega$ for different values of the channel aspect ratio $\epsilon$. Again, one observes that the qualitative trends for different values of $\epsilon$ are the same as those described above; however, the enhanced diffusion (resonance) peak becomes more pronounced as $\epsilon$ increases.  While the saturated value of $v$ is controlled by $\epsilon$, the convergence of $v$ to this value remains independent of $\epsilon$. Since the bottleneck opening of the channel increases with $\epsilon$, both $v$ and $D_{\mathrm{eff}}$ increase monotonically with increasing $\epsilon$. As one would expect, for $\epsilon = 1$, i.e., a flat and spatially symmetric channel, $v$ becomes zero in the limit $\omega \to 0$. 

\section{Concluding Remarks}\label{Sec: Conclusions}

In this work, we have studied the diffusive behavior of gravitactic chiral self-propelled particles confined within a two-dimensional asymmetric channel. Using Brownian dynamics simulations, we have demonstrated the emergence of resonant diffusion when the external torque $\omega$ approaches the intrinsic angular velocity $\omega_{0}$, occurring only for low rotational diffusion rates and does not persist beyond moderate values of $\omega_{0}$. This resonance leads to a pronounced accumulation of particles near the upper-left corner of the channel, resulting in an enhanced peak in the effective diffusion coefficient. The origin of this resonance is traced to an underlying bifurcation of the classical overdamped driven pendulum. We found that, far below the bifurcation ($\omega \ll \omega_{0}$), the particles exhibit rectification in the positive $x$ direction. Above the bifurcation ($\omega > \omega_{0}$), $\omega$ increasingly drives the particles along the horizontal direction and a monotonic increase in the average velocity with $\omega$ is observed. Moreover, we have shown that the transport characteristics, including the average velocity and effective diffusion coefficient, are significantly influenced by the rotational diffusion rate, intrinsic angular velocity, and aspect ratio of the channel. In particular, both the average velocity and effective diffusion coefficient exhibit a peak at an optimal value of the rotational diffusion rate near the bifurcation ($\omega \approx \omega_{0}$). The enhanced diffusion peak becomes more pronounced as the aspect ratio increases, and the average velocity saturates at higher values for channels with wider bottleneck openings. 

These results could stimulate the development of new strategies for controlling the diffusion of asymmetric active particles, such as chiral bacteria (e.g., E. coli and V. cholerae) and artificially designed microswimmers, in entropic ratchet systems. Furthermore, since the external torque in our model can be realized not only via gravity but also through hydrodynamic or magnetic fields, our findings are relevant to a wide range of applications, including microfluidic and laboratory-on-a-chip devices for particle sorting~\cite{Song_et_al@ELPS:2023,Wu_et_al@Chaos:2023,Aristov_et_al@SM:2013}, targeted drug delivery~\cite{Bozuyuk_et_al@ACSN:2018,Gao_et_al@Small:2012}, and the design of advanced active materials~\cite{Mooney_et_al@AM:2024,Sanchez_et_al@ACIE:2015}.

\section{Acknowledgments}

This work was supported by the Jaypee Institute of Information Technology, Noida. The authors gratefully acknowledge the Ramanujan Universe HPC facility at Jaypee Institute of Information Technology, Noida, for providing the computational resources.

\appendix
\section{Choice of parameters}\label{Sec: Appendix}

The findings reported above can also be realized experimentally with gravitactic chiral active particles diffusing in an asymmetric channel, which can be fabricated on a substrate using microprinting~\cite{Mahmud_et_al@NP:2009}. For experimental relevance, it is instructive to have an estimate of parameters in real units. It should be noted that the choice of parameters imposes constraints on physical systems. For thermal noise, the Stokes-Einstein relations hold, and the typical parameter values for active particles with radii $a \sim 100~\mathrm{nm}$, mass $M \sim 10^{-16}~\mathrm{kg}$, and moment of inertia $I \sim 10^{-31}~\mathrm{kg\,m^2}$ in water moving in an asymmetric channel at room temperature are given by ${\rm w}_\mathrm{max} = L \sim 1~\mu\mathrm{m}$, $\epsilon = 0.1$, $V_0 \sim 1\text{--}50~\mu\mathrm{m/s}$, $\Omega_{0} \sim 0\text{--}10^{3}~\mathrm{s^{-1}}$, $\Omega \sim 0\text-1~\mathrm{s^{-1}}$, $\gamma_t \sim 10^{-9}~\mathrm{kg/s}$, $\gamma_r \sim 10^{-22}~\mathrm{kg\,m^2/s}$, $D_t \sim 2 \times 10^{-12}~\mathrm{m^2/s}$, $D_r \sim 10~\mathrm{s^{-1}}$, and $\tau \sim 0.5~\mathrm{s}$. While for athermal noise, we instead take $D_t \sim 10^{-11}-10^{-9}~\mathrm{m^2/s}$ and $D_r \sim 10~\mathrm{s^{-1}}$. These values were utilized to determine the dimensionless parameters used in the text.    

\bibliography{APS} 



\end{document}